\documentclass[11pt]{article}
\usepackage{a4wide}
\usepackage{graphics,epsfig,epsf,psfrag}

\usepackage{amsmath,amssymb,mathrsfs,latexsym}
\usepackage{slashed}
\usepackage{bbm}
\usepackage[usenames]{color}

\newcommand{\non}{\nonumber}

\newcommand{\Amp}{\mathcal{A}}

\newcommand{\lang}{\left\langle}
\newcommand{\rang}{\right\rangle}

\begin{document}
\title{Confirming the molecular nature\\ of the $Z_b(10610)$ and the $Z_b(10650)$}
\author{Martin Cleven$^{1,}$\thanks{{\it Email address:} m.cleven@fz-juelich.de},
        Qian Wang$^{1,}$\thanks{{\it Email address:} q.wang@fz-juelich.de},
        Feng-Kun Guo$^{2,}$\thanks{{\it Email address:} fkguo@hiskp.uni-bonn.de},
        Christoph Hanhart$^{1,3,}$\thanks{{\it Email address:}
        c.hanhart@fz-juelich.de}, \\
        Ulf-G. Mei{\ss}ner$^{1,2,3,}$\thanks{{\it Email address:} meissner@hiskp.uni-bonn.de},
        and Qiang Zhao$^{4,}$\thanks{{\it Email address:}
        zhaoq@ihep.ac.cn}\\[3mm]
  {\small
       $^1$\it Institut f\"{u}r Kernphysik and J\"ulich Center for Hadron
          Physics},\\
          {\small \it Forschungszentrum J\"{u}lich, D--52425 J\"{u}lich, Germany}\\
       {\small $^2$\it Helmholtz-Institut f\"ur Strahlen- und Kernphysik and
          Bethe Center for Theoretical Physics,}\\
       {\small \it  Universit\"at Bonn, D--53115 Bonn, Germany}\\
       {\small$^3$\it Institute for Advanced Simulation,
          Forschungszentrum J\"{u}lich, D--52425 J\"{u}lich, Germany}\\
       {\small $^4$\it Institute of High Energy Physics and Theoretical Physics Center for Science Facilities,}\\
       {\small \it Chinese Academy of Sciences, Beijing 100049, China}
          }
\date{\today}

\maketitle
\begin{abstract}
  The decays of the $Z_b(10610)$ and the $Z_b(10650)$ to $\Upsilon(nS)\pi$,
  $h_b(mP)\pi$ and $\chi_{bJ}(mP)\gamma$ ($n=1,2,3$, $m=1,2$ and $J=0,1,2$)
  are investigated within a nonrelativistic effective field theory. It is
  argued that, while the decays to $\Upsilon(nS) \pi$ suffer from potentially
  large higher order corrections, the $P$-wave transitions of the $Z_b$ states
    are dominated by a single one loop diagram and therefore
  offer the best possibility to confirm the nature of the $Z_b$ states as
  molecular states and to further study their properties. We give nontrivial
  and parameter-free predictions for the ratios of various partial
  widths of the $Z_b$ and $Z_b'$ into final states with $h_b(mP)\pi$ and $\chi_{bJ}(mP)\gamma$.
  While such relations appear naturally in the molecular picture for the
  mentioned transitions, they are not expected to hold for any other scenario.
  In addition, the branching fractions for the neutral $Z_b$-states to $\chi_{bJ}\gamma$ are
  predicted to be of order $10^{-4}$--$10^{-3}$. This provides a fine test of
  the molecular nature in future high-luminosity experiments.
\end{abstract}


\newpage

\section{Introduction}
\label{sec:introduction}

Recently the Belle Collaboration found two narrow structures, namely the
$Z_b(10610)=Z_b^{\pm}$ and the $Z_b'(10650)=Z_b^{\prime\pm}$, in the
$\Upsilon(nS)\pi^{\pm}$ $(n=1,2,3)$ and $h_b(mP)\pi^{\pm}$ $(m=1,2)$ invariant
masses of the $\Upsilon(5S)\to \Upsilon(nS)\pi^+\pi^-$ and $\Upsilon(5S)\to
h_b(mP)\pi^+\pi^-$ decays processes~\cite{Belle:2011aa}. In their latest data,
also the open-bottom channels $B^*\bar B$ and $B^*\bar B^*$ are
seen~\cite{Adachi:2012cx}. The fact that they lie in the bottomonium mass region
and they are charged means that they cannot be the conventional bottomonium
mesons, and their isospins are 1. The observation of the neutral state with a
mass consistent with that of the $Z_b(10610)$~\cite{Adachi:2012im} presents a
further confirmation that they are members of  isotriplets. If they exist as
observed by the Belle Collaboration, they must be exotic states with a pair of
hidden $b\bar{b}$ and valence light quarks. Considering parity and charge
parity, their quantum numbers should be $I^G(J^{P})=1^+(1^{+})$, and the charge
parity of the neutral state is negative.

Assuming that the total width of $Z_b^{(\prime)}$ is saturated by the seven
channels already observed experimentally, i.e. $\Upsilon(nS)\pi$ $(n=1,2,3)$,
${h_b(mP)\pi\,(m=1,2)}$, $B\bar{B}^*+B^*\bar B$ and $B^*\bar{B}^*$, Belle gives
the branching ratio of each channel in $\Upsilon(5S)$ three--body
decays~\cite{Adachi:2012cx}. The proximity of  the states to the $B^{(*)}\bar
B^*$ thresholds leads to the suggestion that they could be hadronic molecules of
the corresponding
states~\cite{Bondar:2011ev,Cleven:2011gp,Nieves:2011zz,Zhang:2011jja,Yang:2011rp,Sun:2011uh,Ohkoda:2011vj,Li:2012wf,Ke:2012gm},
to be distinguished from the compact $\bar b\bar q b q$
tetraquarks~\cite{Li:2012wf,Guo:2011gu,Ali:2011ug}. By hadronic molecules, we
mean states composed of hadrons --- they can be bound states (poles on the
physical sheet with respect to the bottom-meson channel with a mass smaller than
the threshold value), resonances or virtual states (both on the second sheet
with respect to the relevant bottom-meson channel --- the former above, the
latter below the threshold). Based on a nonrelativistic effective field theory
(NREFT)~\cite{Guo:2009wr,Guo:2010ak}, the authors of Ref.~\cite{Cleven:2011gp}
show that the $h_b(1P,2P)\pi^\pm$ data can be described within the bound state
scenario. Therein, the $Z_b^{(\prime)}B^{(*)}\bar B^*$ coupling constants are
related to the binding energies using a model-independent relation for $S$-wave
shallow bound states~\cite{Weinberg:1965zz,Baru:2003qq}. However, since now data
on the decays of  $Z_b^{(\prime)}$ to open bottom channels are available, we can
chose a more general ansatz and take these couplings from data directly --- in
this way our results are valid for bound states, resonances as well as virtual
states. Since we start from the assumption that the
 $Z_b$ states are purely molecular states, their decays into the $\Upsilon(nS)\pi$ and $h_b(mP)\pi$
 can only happen via
$B \bar{B}^*+B^*\bar B$ (for simplicity, $B \bar{B}^*$ will be used to represent
$B \bar{B}^*+B^*\bar B$ in the following) and $B^*\bar{B}^*$ loops. Since both
$Z_b$'s are located very close to the corresponding open-bottom threshold, the
system can in principle be examined by the NREFT approach.

As shown in Refs.~\cite{Guo:2009wr,Guo:2010ak}, a systematic power
counting can be established for the $\pi$ or $\eta$ emissions
between charmonium states. Because the $S$-wave and $P$-wave heavy
quarkonia couple to the open-flavor heavy meson and anti-meson in a
$P$-wave and an $S$-wave, respectively, the transitions studied in
Ref.~\cite{Guo:2010ak} can be classified into three groups, namely
transitions between the $S$-wave heavy quarkonium states, $P$-wave
states, and between the $P$- and $S$-wave states. The decay
amplitudes of different groups have their own nonrelativistic
velocity counting. Because the $Z_b$ states have positive parity,
they couple to the bottom and anti-bottom mesons in an $S$-wave.
Thus, the transitions $Z_b\to \Upsilon(nS)\pi$ are analogous to
those between $P$- and $S$-wave quarkonia, while the $Z_b\to
h_b(mP)\pi$ processes are similar to those between two $P$-wave
quarkonia. The main difference is that the normal heavy quarkonium
transitions with the emission of a pion studied in
Ref.~\cite{Guo:2010ak} break isospin symmetry while the $Z_b$ decays
do not. Thus, by studying the $Z_b \ (Z_b')\to \Upsilon(nS)\pi$ and
$h_b(mP)\pi$, we can examine the power counting rules established in
Ref.~\cite{Guo:2010ak}, and also better understand the properties of
these two exotic states.

Further insight can be gained by studying radiative decays of the neutral $Z_b$
states into bottomonia. Because the $\chi_{bJ}\,(J=0,1,2)$ states are the spin
partners of the $h_b$ of the same principal quantum number, they couple to the
bottom mesons with the same coupling constant in the heavy quark limit. Thus,
since molecular states can decay via bottom meson loops only, the radiative
decays $Z_b^{(\prime)0}\to \chi_{bJ}(nP)\gamma$ are related to the pionic decays
$Z_b^{(\prime)\pm}\to h_b(nP)\pi^\pm$. Note that these transitions would be
unrelated if the $Z_b$ states were of tetraquark nature. Similar considerations
were made in Ref.~\cite{Guo:2011dv} for hindered M1 transitions of the $P$-wave
charmonia. These decay channels have not been observed so far, but they could be
potentially important in confirming the molecular nature of the $Z_b$ states,
and thus are worthwhile to study experimentally.

In this paper, we will assume that the $Z_b^{(\prime)}$ are dynamically
generated  from the $B^{(*)}\bar B^*$ interactions, i.e. hadronic molecules of
the $B^{(*)}\bar B^*$. We will try to identify the quantities which are
sensitive to such a scenario. Section~\ref{sec:nonrel} contributes to the power
counting in the NREFT framework of the relevant decays,  which are the decays of
the $Z_b$ states into the $\Upsilon\pi$, $h_b\pi$, $\chi_{bJ}\gamma$ --- in
particular we show that not all decays are accessible to the formalism. The
numerical results are given in Sec.~\ref{sec:results}. In
Sec.~\ref{sec:comparison}, we compare our results to previous calculations and
make some comments. A brief summary is presented in Sec.~\ref{sec:summary}. The
loop function used in the calculations and the decay amplitudes in the NREFT are
collected in Appendix~\ref{app:calnonrel}. As a cross check, we also calculate
the same quantities using a Lorentz covariant formalism, and the formulas are
summarized in Appendix~\ref{app:calrel}.

\section{Power Counting}\label{sec:nonrel}

In Ref.~\cite{Guo:2009wr}, a NREFT method was introduced to study the meson loop
effects in the heavy quarkonium transitions. The power counting scheme was
analyzed in detail in Ref.~\cite{Guo:2010ak}. The key quantities here are the
velocities of the intermediate mesons. In this section, we briefly review the
ideas of Refs.~\cite{Guo:2009wr,Guo:2010ak} together with an improved discussion
for the higher loop diagrams.

In general, the heavy meson velocities relevant for the decay of
some particle $X$ may be estimated as
$v_X\sim\sqrt{|{M_X}-2m_B|/m_B}$, where the absolute value indicates
that the formula can be used for both bound systems as well as
resonances. The analogous formula holds when the two heavy mesons
merge to a quarkonium in the final state. According to the rules of
a nonrelativistic effective field theory~\cite{david} (for a review,
see e.g.~\cite{Brambilla:2004jw}), the momentum and non-relativistic
energy count as $v_X$ and $v_X^2$, respectively. For the integral
measure one finds $v_X^5/(4\pi)^2$. The heavy meson propagator
counts as $1/v_X^2$. The leading order $S$-wave vertices do not have
any velocity dependence, while the case for the $P$-wave vertices is
more complicated: it scales either as $v_X$ when the momentum due to
$P$-wave coupling contracts with another internal momentum, or as
the external momentum $q$ when $q$ is contracted.

\begin{figure}
\centering
\includegraphics[width=\linewidth]{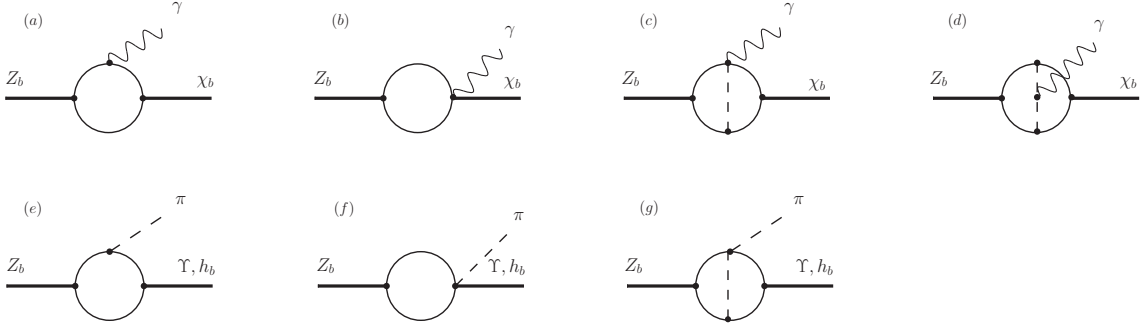}
\caption{Schematic one- and two-loop diagrams of the transitions $Z_b\to \Upsilon\pi,~h_b\pi$ and $\chi_{bJ}\gamma$.}
\label{fig:PowerCounting1}
\end{figure}
We start with the radiative transitions as shown in the upper row of
Fig.~\ref{fig:PowerCounting1}. If the $Z_b$ states are molecular
states, their spin wave functions contain both  $s_{b\bar b}=0$ and
1 components~\cite{Bondar:2011ev}, where $s_{b\bar b}$ is the total
spin of the $b\bar b$ component. Thus, the radiative decays of the
$Z_b$ states into the spin-triplet $\chi_{bJ}$ can occur without
heavy quark spin flip and survive in the heavy quark limit. This is
different from the M1 transitions between two $P$-wave heavy
quarkonia which have been analyzed in Ref.~\cite{Guo:2011dv}. Both
the couplings of $Z_b$ and $\chi_{bJ}$ to a pair of heavy mesons are
in an $S-$wave, and the photon coupling to the bottom mesons is
proportional to the photon energy $E_\gamma$. For the diagram of
Fig.~\ref{fig:PowerCounting1}~(a) the amplitude therefore scales as
\begin{equation}
 \frac{\bar v^5}{(4\pi)^2}\frac{1}{(\bar v^2)^3}E_\gamma\sim
 \frac{E_\gamma}{(4\pi)^2 \bar v} \ ,
 \label{eq:Apower}
\end{equation}
where  the velocity that appears is $\bar v=(v_Z+v_\chi)/2\simeq
v_\chi/2$~\cite{Guo:2012tg}, since $v_Z\sim v_{Z'}\simeq 0.02$, if the central
values of the measured $Z_b^{(\prime)}$ masses, 10607.2~MeV and 10652.2~MeV, are
used, while $v_\chi$ ranges from 0.12 for the $\chi_{bJ}(3P)$ to 0.26 for the
$\chi_{bJ}(2P)$ to 0.37 for the $\chi_{bJ}(1P)$. Here we used the mass of the
$\chi_{bJ}(3P)$, 10.53~GeV, as reported by the ATLAS
Collaboration~\cite{Aad:2011ih}. In the following, we will count $\bar v$ as
$\mathcal{O}(v_\chi)$. The scaling ensures that the amplitude gets larger when the bottomonium in the
final state is closer to the open-bottom threshold. Thus, we expect for the
absolute value of the decay amplitude from this diagram
\begin{eqnarray}
\left|\frac{\mathcal{A}_{\chi_{bJ}(1P)\gamma}}{E_\gamma}\right|:
\left|\frac{\mathcal{A}_{\chi_{bJ}(2P)\gamma}}{E_\gamma}\right|:\left|\frac{\mathcal{A}_{\chi_{bJ}(3P)\gamma}}{E_\gamma}\right|
\sim \frac1{v_{1P}} : \frac1{v_{2P}} : \frac1{v_{3P}}
=  1:1.4:3.1,
\label{eq:Aratio}
\end{eqnarray}
if the $\chi_{bJ}(nP)B\bar B$ coupling constants take the same value. Diagram
(a) can be controlled easily in theory. Thus, clear predictions can be made
whenever diagram (a) dominates. In the following we will identify such dominant
decays based on the power counting for the NREFT.

As for Fig.~\ref{fig:PowerCounting1}~(b), the coupling $\chi_b B
\bar B \gamma$ cannot be deduced by gauging the coupling of
$\chi_{bJ}$ to a $B\bar B$-meson pair. Thus, it has to be gauge
invariant by itself and proportional to the electromagnetic field
strength tensor $F^{\mu\nu}$. This gives a factor of photon energy
$E_\gamma$ and the amplitude of Fig.~\ref{fig:PowerCounting1}~(b)
scales as
\begin{equation}
\label{eq:leadrad}
 \frac{v_Z^5}{(4\pi)^2}\frac{1}{(v_Z^2)^2} E_\gamma\sim
 \frac{E_\gamma v_Z}{(4\pi)^2} \ ,
\end{equation}
where we have assumed that the corresponding coupling is of natural size. Thus,
diagram (b) is suppressed compared to diagram (a) at least by a factor of $v_Z
v_\chi<0.01$ for the decay to $\chi_{bJ}(1P)$ and even smaller for the excited
states.

The situation is more complicated for the graph displayed
in Fig.~\ref{fig:PowerCounting1}~(c). Here we
have a two-loop diagram, so that the velocities running in different loops are
significantly different --- the one in the loop connected to the $Z_b$ is $v_Z$,
and the other is $v_\chi$. It is important to count them separately since
$v_Z\ll v_\chi$ due to the very close proximity of the $Z_b$ to the
threshold\footnote{The concept applied here is analogous to the scheme by now
well established for the effective field theory for reactions of the type $NN\to
NN\pi$ --- see Ref.~\cite{report} for a review.}. The internal pion momentum
scales as the larger loop momentum, and thus the pion propagator should be
$\sim1/(m_B^2 v_\chi^2)$. This leaves us with
\begin{equation}
 \frac{v_Z^5}{(4\pi)^2}\frac{1}{(v_Z^2)^2}\frac{v_\chi^5}{(4\pi)^2}
 \frac{1}{(v_\chi^2)^2}\frac{1}{m_B^2v_\chi^2}\frac{E_\gamma g}{F_\pi}
 \frac{g}{F_\pi}m_B^4 \sim \frac{v_Z}{v_\chi}
 \frac{E_\gamma g^2 m_B^2}{(4\pi)^2\Lambda_\chi^2}~,
\end{equation}
where $F_\pi$ is the pion decay constant in the chiral limit, the
factor $m_B^4$ has been introduced to give the same dimension as the
estimate for the first two diagrams, and the hadronic scale was
introduced via the identification $\Lambda_\chi=4\pi F_\pi$. Thus
the two-loop diagram is suppressed compared to the leading one,
Eq.~(\ref{eq:leadrad}), by a factor $v_Z g^2
m_B^2/\Lambda_\chi^2\sim 0.1$, where we used for the coupling
$B^*\to B\pi$ the value $g=0.5$ (a recent lattice calculation gives
$0.449\pm0.051$~\cite{Detmold:2011bp}), and $\Lambda_\chi\sim1$~GeV.
 It can easily
be seen that Fig.~\ref{fig:PowerCounting1}~(d) gives the same
contribution which also reflects the fact that they are both
required at the same order to ensure gauge invariance. Thus, from
our power counting it follows that the loop diagrams of
Fig.~\ref{fig:PowerCounting1} (b)-(d) provide a correction of at
most 10\%. We will therefore only calculate diagram (a) explicitly
and introduce a 10\%  uncertainty for the amplitudes which
corresponds to 20\% for the branching ratios. Higher loop
contributions are to be discussed later.

Next we consider the hadronic transition $Z_b\to h_b\pi$.  This decay
has already been studied in Ref.~\cite{Cleven:2011gp} in the same
formalism. Again, since the $h_b$ has even parity, its coupling to
the bottom mesons is in an $S$-wave. In addition, the final state
must be in a $P$-wave to conserve parity, such that the amplitude
must be linear in the momentum of the outgoing pion, $q$. We therefore find
for the one-loop contribution of Fig.~\ref{fig:PowerCounting1} (e)
\begin{equation}
\frac{\bar v^5}{(4\pi)^2}\frac{1}{(\bar  v^2)^3} \frac{gq}{F_\pi}
\sim g\frac{q F_\pi}{\bar v \Lambda_\chi^2} \ ,\label{eq:hbpia}
\end{equation}
while Fig.~\ref{fig:PowerCounting1} (f)  gives
\begin{equation}
 \frac{v_Z^5}{(4\pi)^2}\frac{1}{(v_Z^2)^2}\frac{q}{F_\pi}\sim \frac{v_Z q
   F_\pi}{\Lambda_\chi^2}.
\end{equation}
Notice that the pion has to be emitted after the loop if the $Z_b$ is a pure
hadronic molecule, so that the velocity in the counting should be $v_Z$ instead
of $v_h$. This is suppressed by $v_hv_Z/g$ which leads to a correction of the
order of  2\% noticing that $v_h\simeq v_\chi$.

The two-loop diagram Fig.~\ref{fig:PowerCounting1} (g) contributes as
\begin{equation}
\frac{v_Z^5}{(4\pi)^2}\frac{1}{(v_Z^2)^2}\frac{v_\chi^5}{(4\pi)^2}\frac{1}{(v_\chi^2)^2}\frac{1}{m_B^2v_\chi^2}\frac{gq}{F_\pi}
  \frac{E_\pi }{F_\pi^2}m_B^3
=\frac{v_Z}{v_h}\frac{ g F_\pi q^2 m_B}{\Lambda_\chi^4} \ ,
\label{eq:hbpi_g}
\end{equation}
where we have used that the energy from the $\pi B\to \pi B$ vertex can be
identified with the energy of the outgoing pion~\cite{Lensky:2005jc}, $E_\pi\sim
q$.  The $B^*B\pi$ vertex contributes a factor of the external momentum $q$
since the $Z_b\to h_b\pi$ is a $P$-wave decay, and this is the only $P$-wave
vertex. Therefore, this diagram is suppressed compared to the leading loop,
Eq.~\eqref{eq:hbpia}, by a factor $v_Z m_B q/\Lambda_\chi^2$ which is smaller
than 10\%. Thus,  also for the transitions $Z_b^{(\prime)}h_b\pi$ we may only
calculate the leading one-loop diagrams, Fig.~\ref{fig:PowerCounting1} (e), and
assign an uncertainty of 10\% to the rates which gives an uncertainty of 20\%
for the branching ratios. Higher loop contributions are to be discussed later.

Finally, we consider at the decay channel $Z_b\to \Upsilon \pi$. Here the final
state is in an $S$-wave, but the coupling of the $\Upsilon$ to $\bar BB$ is in a
$P$-wave. For diagram (e) the momentum due to this coupling has to scale as the
external pion momentum. Together  with the pionic coupling that is also linear
in the pion momentum, the amplitude is thus proportional to $q^2$. The one-loop
diagram for $Z_b\to \Upsilon\pi$ via Fig.~\ref{fig:PowerCounting1} (e), is
therefore estimated as
\begin{equation}
 \frac{\bar v^5}{(4\pi)^2}\frac{q}{(\bar v^2)^3}\frac{gq}{F_\pi} \sim g\frac{q^2
   F_\pi}{\bar v\Lambda_\chi^2}.
   \label{eq:upsilonpi}
\end{equation}
The diagram Fig.~\ref{fig:PowerCounting1} (f) on the other hand gives
\begin{equation}
 \frac{v_Z^5}{(4\pi)^2}\frac{1}{(v_Z^2)^2}\frac{E_\pi}{F_\pi} m_B
\sim
\frac{v_Z E_\pi F_\pi m_B}{\Lambda_\chi^2},
\end{equation}
where $m_B$ is introduced in order to get the same dimension as
Eq.~\eqref{eq:upsilonpi}. Compared to the one-loop diagram (e) this
is a relative suppression of order $v_\Upsilon v_Z m_B/q$ which is
less than 10\% for all the $\Upsilon$ states,  where the value of
$v_\Upsilon$ is about 0.46, 0.33 and 0.22 for the $1S$, $2S$ and
$3S$ states, respectively. The two-loop contribution with the
exchange of a pion, Fig.~\ref{fig:PowerCounting1} (g), is estimated
as
\begin{equation}
  \frac{v_Z^5}{(4\pi)^2}\frac{1}{(v_Z^2)^2}\frac{v_\Upsilon^5}{(4\pi)^2}
\frac{1}{(v_\Upsilon^2)^2}\frac{1}{m_B^2v_\Upsilon^2}\frac{v_\Upsilon^2
g}{F_\pi} \frac{E_\pi }{F_\pi^2}m_B^5 \sim \frac{v_Zv_\Upsilon
gE_\pi F_\pi}{\Lambda_\chi^4}m_B^3 \ .
\end{equation}
Thus, the strength of two-loop diagram relative to leading one-loop diagram
 is estimated  for the $\Upsilon\pi$ as
$ v_Z v_\Upsilon^2 m_B^3/(\Lambda_\chi^2 q)$. Numerically, this corresponds to a
factor of around 0.6 for the $\Upsilon(1S,2S)\pi$ and 0.7 for the
$\Upsilon(3S)\pi$ amplitudes. As a consequence, the branching ratios for these
transitions can only be calculated with rather large uncertainties up to 100\%.

\begin{figure}
\centering
\includegraphics[width=0.5\linewidth]{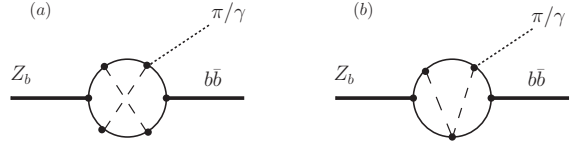}
\caption{Two three-loop diagrams contributing to the decays of the $Z_b$ into a heavy quarkonium and a pion or photon.}
\label{fig:PowerCounting2}
\end{figure}

The heavy meson velocities relevant for the mentioned transitions
range from 0.02 to 0.5 --- in momenta this is a range from 0.1 to
2.5~GeV. While pion contributions are expected to be suppressed
significantly and can be controlled within chiral perturbation
theory for pion momenta of up to 500 MeV and smaller, higher pion
loop contributions might get significant for momenta beyond 1~GeV.
We will now study those higher pion loops within the power counting
scheme outlined above. We will start with the three-loop diagrams,
as shown in Fig.~\ref{fig:PowerCounting2} considering first diagram~(a).
The results can be easily generalized to higher loops as shown
below. Compared to the two-loop diagrams in
Fig.~\ref{fig:PowerCounting1}, there are one more pion propagator
and two more bottom meson propagators in the transition and none of
them is connected to the external heavy quarkonia. In addition,
there is no two-bottom-meson unitary cut present. As a consequence,
we have to use a relativistic power counting --- c.f.
Ref.~\cite{david}. Then pion momentum and energy are of order $m_B
v_{b\bar b}$, with $v_{b\bar b}$ the velocity of the $B$-meson
connected to the $b\bar b$-meson in the final state. Both the
energies and momenta of the additional bottom mesons are now of the
same order, such that the bottom-meson propagator is counted as
$1/v_{b\bar b}$. The integral measure reads $v_{b\bar
b}^4/(4\pi)^2$. Therefore, the additional factor as compared to the
two-loop diagrams  is
\begin{equation}
\label{higherloop1}
\frac{v_{b\bar b}^4}{(4\pi)^2} \frac1{v_{b\bar b}^4} \frac{\left(g v_{b\bar b}\right)^2}{F_\pi^2}m_B^2
= \left( \frac{g\,m_B v_{b\bar b}}{\Lambda_\chi} \right)^2 \ .
\end{equation}
If $v_{b\bar b}\sim 0.4$, then the three-loop diagram (a) is of
similar size as that of the two-loop diagrams.  This is the case for
the processes with the $h_b(1P), \chi_{bJ}(1P)$, and
$\Upsilon(1S,2S)$. For smaller values of $v_{b\bar b}$, it is
suppressed. In diagram (b) there is only one more bottom meson
propagator and the additional $B\pi B\pi$ vertex is in an $S$-wave.
We obtain
\begin{equation}
\label{higherloop2}
\frac{v_{b\bar b}^4}{(4\pi)^2} \frac1{v_{b\bar b}^3} \frac{v_{b\bar b}}{F_\pi^2}m_B^2
= \left( \frac{m_B v_{b\bar b}}{\Lambda_\chi} \right)^2.
\end{equation}
Four and higher loop diagrams that cannot be absorbed by using
physical parameters may now be estimated by applying a proper number
of factors of the kind of Eqs. (\ref{higherloop1}) and
(\ref{higherloop2}). It is easy to see that also additional
topologies provide analogous factors. Since $m_B/\Lambda_\chi\sim
5$, higher loops get increasingly important, if $v_{b\bar b}>0.2$.
For $v_{b\bar b}\sim0.2$, the three and more loops are of  the same
order as the two-loop diagrams, which is the case for the $2P$
states, and thus suppressed in comparison with the one-loop
contribution. For the $3P$ bottomonia in the final state, the value
of $v_{b\bar b}$ is even smaller, and the multiple loops are even
suppressed as compared to the two-loop contribution.

To summarize the findings of the power counting analysis, we conclude that the
calculation  of one-loop triangle diagrams as depicted in
Fig.~\ref{fig:PowerCounting1} (a) and (e) is a good approximation, with a
controlled uncertainty, to the transitions of the $Z_b^{(\prime)}$ into the
$\chi_{bJ}(2P,3P)\gamma$ and $h_{b}(2P)\pi$ --- for the $h_b(3P)\pi$ the phase
space is too limited. But similar calculations are not applicable to the decays
into the $1P$ states as well as the $\Upsilon(nS)\pi$. For the decays $Z_b\to
\Upsilon(nS)\pi$, the contribution from the two-loop diagrams is not suppressed
or even enhanced compared to the one-loop contribution. In light of this
discussion, it becomes clear why the pattern of branching fractions for these
channels (see Tab.~\ref{table-Belle-branching-ratios}) $\mathcal
B[\Upsilon(2S)\pi]>\mathcal B[\Upsilon(3S)\pi]\gg\mathcal B[\Upsilon(1S)\pi]$
cannot be reproduced by calculating the three-point diagrams in the NREFT
formalism, which always favors the $1S$ to the $2S$ transition and the $2S$
compared to the $3S$ transition due to the factor $q^2$ in
Eq.~\eqref{eq:upsilonpi}.

\begin{table}[t]
\centering
\scriptsize
\caption{Left: Preliminary measurements of the branching ratios for $Z_b^{(\prime)}$ from
the Belle Collaboration~\cite{Adachi:2012cx}. Right:
Masses of the various particles used here~\cite{Adachi:2011ji,Behringer:2012ab}.}
\medskip
\renewcommand{\arraystretch}{1.3}
\begin{tabular}{|l|c|c|}
\hline
Branching ratio ($\%$) & $Z_b(10610)$ & $Z'_b(10650)$ \\ \hline\hline
$\Upsilon(1S)\pi^+$ & $ 0.32\pm0.09$ & $ 0.24\pm0.07$ \\ \hline
$\Upsilon(2S)\pi^+$ & $ 4.38\pm1.21$ & $ 2.40\pm0.63$ \\ \hline
$\Upsilon(3S)\pi^+$ & $ 2.15\pm0.56$ & $ 1.64\pm0.40$ \\ \hline
$h_b(1P)\pi^+$ & $ 2.81\pm1.10$ & $ 7.43\pm2.70$ \\ \hline
$h_b(2P)\pi^+$ & $ 4.34\pm2.07$ & $ 14.82\pm6.22$ \\ \hline
$B^+\bar B^{*0}+\bar B^0 B^{*+}$ & $ 86.0\pm3.6$ & $-$ \\ \hline
$B^{*+}\bar B^{*0}$ & $-$ & $ 73.4\pm7.0$ \\ \hline\hline
\end{tabular}
\hspace{0.4cm}
\begin{tabular}{|l|c|l|c|}
\hline
  & Mass [GeV] & & Mass [GeV] \\ \hline\hline
$\Upsilon(1S)$ & $\phantom{0}9.460$ & $\chi_{b0}(1P)$ & \phantom{0}9.859 \\ \hline
$\Upsilon(2S)$ & $ 10.023$ & $\chi_{b1}(1P)$ & \phantom{0}9.893 \\ \hline
$\Upsilon(3S)$ & $ 10.355$ & $\chi_{b2}(1P)$ & \phantom{0}9.912 \\ \hline
$h_b(1P)$ & $ \phantom{0}9.899$ & $\chi_{b0}(2P)$ & 10.233 \\ \hline
$h_b(2P)$ & $10.260$ & $\chi_{b1}(2P)$ & 10.255 \\ \hline
$B$ & $\phantom{0}5.279$ & $\chi_{b2}(2P)$ & 10.269 \\ \hline
$B^{*}$ & $\phantom{0}5.325$ & $\pi$ & \phantom{0}0.138 \\ \hline\hline
\end{tabular}
\label{table-Belle-branching-ratios}
\end{table}

\section{Results}\label{sec:results}
In this section, we investigate quantitatively the decays of the $Z_b$ states
through  heavy meson loops. The coupling of these states to the heavy meson
fields $H_a=\vec V_a\cdot \sigma+P_a$ and $\bar H_a=-\vec{\bar V}_a\cdot
\vec{\sigma}+\bar P_a$ with $V_a$ ($\bar V_a$) and $P_a$ ($\bar P_a$)
annihilating the vector and pseudoscalar (anti-)heavy mesons, respectively, is
given by the Lagrangian
\begin{equation}\nonumber
 \mathcal{L}_Z = i \frac{z}{2} \lang Z^{\dag i}_{ba} H_a \sigma^i{\bar
H}_b\rang + {\rm H.c.}, \label{eq:Lz}
\end{equation}
where the  $Z_b$ states are given by a $2\times 2$ matrix
\begin{equation}\nonumber
 Z^i_{ba} = \left(
         \begin{array}{cc}
           \frac1{\sqrt{2}} Z^{0i} & Z^{+i} \\
           Z^{-i} & - \frac1{\sqrt{2}} Z^{0i} \\
         \end{array}
       \right)_{ba}.
\end{equation}
We incorporate the experimental observation that the $Z_b^{(\prime)}$
couples only to $B^{(*)}\bar B^*$ via the Lagrangian,
\begin{equation}\nonumber
 \mathcal{L}_{Z,Z'}=z'\varepsilon^{ijk}\bar V^{\dagger i} Z^j V^{\dagger
   k}+z\left[\bar V^{\dagger i}Z^iP^\dagger-\bar P^\dagger Z^iV^{\dagger
     i}\right]+ {\rm H.c.}
\end{equation}
Fitting to the experimental data we find
\begin{equation}
 z=(0.79\pm0.05)~{\rm GeV}^{-1/2}, \qquad z'=(0.62\pm0.07)~{\rm GeV}^{-1/2}, \label{eq:couplings-NREFT}
\end{equation}
and especially
\begin{equation}
 \frac{z}{z'}= 1.27 \pm0.16 \  \label{eq:coupling-ratio}
\end{equation}
which deviates from unity by 2$\sigma$. This deviation indicates a significant
amount of spin symmetry violation, which, however, is not unnatural for
very-near-threshold states where small differences in masses may imply huge
differences in binding energies resulting in significantly different effective
couplings, as already discussed in Ref.~\cite{Cleven:2011gp}. The
nonrelativistic Lagrangian for the $\chi_{bJ}$ coupling to a pair of heavy
mesons can be found in Ref.~\cite{Fleming:2008yn}, and  the one for the magnetic
coupling of heavy mesons in Ref.~\cite{Hu:2005gf}.

With the amplitudes given in Appendix~\ref{app:calnonrel}, it is straightforward
to  calculate the decay widths of the $Z_b^{(\prime)}\to h_b(mP)\pi$, which are
proportional to $g_1^2$, where $g_1$ is the $P$-wave bottomonium--bottom meson
coupling constant. In the ratio defined as
\begin{equation}
 \xi_m:=\frac{\Gamma[Z'_b\to h_b(mP)\pi]}{\Gamma[Z_b\to h_b(mP)\pi]},
\label{eq:ratio-hb}
\end{equation}
$g_1$ is cancelled out. With the meson masses listed in
Tab.~\ref{table-Belle-branching-ratios}, we obtain
\begin{eqnarray}
 \xi_1=1.21\left|\frac{z'}{z}\right|^2= 0.75,\qquad
 \xi_2=(1.53\pm0.43)\left|\frac{z'}{z}\right|^2= 0.95\pm0.36.
 \label{eq:ratios}
\end{eqnarray}
where the first error in the second term is the theoretical uncertainty due to
neglecting higher order contributions (see the discussion in the previous section),
and the second one also includes the uncertainty of $z'/z$ added in quadrature.
Due to the theoretically uncontrollable higher order contributions for the
decays into the $1P$ states, no uncertainty is given for $\xi_1$.

\begin{table}[t]
\centering
\caption{The ratios $\xi$ of different decay modes in both NREFT and a relativistic framework
  as compared with the experimental data. The NREFT values quoted without uncertainties may be understood
  as order-of-magnitude estimates.}
\medskip
\begin{tabular}{|cccc|}
  \hline\hline
  $\xi$ & NREFT & Rel. & Exp. \\\hline
  $\Upsilon(1S)\pi$ & $0.7$ & 0.7 & $0.47\pm 0.22$ \\
  $\Upsilon(2S)\pi$ & $0.9$ & 0.8 & $0.34\pm 0.15$ \\
  $\Upsilon(3S)\pi$ & $2\pm 2$ & 1.6 & $0.48\pm 0.20$ \\
  $h_b(1P)\pi$ & $0.8$ & 0.7 & $1.65\pm 0.96$ \\
  $h_b(2P)\pi$ & $1.0\pm0.4$ & 0.9 & $2.13\pm 1.44$ \\
  \hline\hline
\end{tabular}
\label{tab:xi}
\end{table}
The predictions are consistent with their experimental counterparts
(see Tab.~\ref{table-Belle-branching-ratios})
\begin{eqnarray}
 \xi_1^{\rm Exp}=1.65\pm0.96, \quad
 \xi_2^{\rm Exp}=2.13\pm1.44.
\end{eqnarray}
Here, new measurements with significantly reduced uncertainties
would be very desirable. A collection of ratios for the decays of
the $Z_b$ states to $h_b(mP)\pi$ and $\Upsilon(nS)\pi$,
respectively, is presented in Tab.~\ref{tab:xi}. The uncertainties,
whenever they are under control theoretically, are also included.
The significant deviations for the $\Upsilon(nS)\pi$ results from
the experimental numbers appear natural, given that for those
transitions higher loops were argued to be at least as important as
the one-loop diagram included here, as outlined in detail in the
previous section.  As a cross check of our nonrelativistic
treatment, we also calculated the same quantities using a Lorentz
covariant formalism with relativistic propagators for all the
intermediate mesons (the formalism is summarized in
App.~\ref{app:calrel}). The results are denoted as `Rel.' in
Tab.~\ref{tab:xi}. They should be compared with the central values
of the NREFT results. The difference reflects relativistic
corrections to the nonrelativistic treatment of the bottom-meson
propagators. In our case, the difference between relativistic and
NREFT calculations here and in the following never exceeds 15\% for
the rates and thus is well below the uncertainty due to higher
loops.

It is important to ask to what extent the above predictions can be
used to probe the nature of the $Z_b$ states. The $Z_b$ and $Z_b'$
are away from the $B\bar B^*$ and $B^*\bar B^*$ thresholds by
similar distances, $m_{Z_b}-m_B-m_{B^*}\simeq m_{Z'_b}-2m_{B^*}$.
Additionally, due to the heavy quark spin symmetry, one may expect
that the loops contribute similarly to the decays of these two
states into the same final state. This is indeed the case. We find
that the ratios aside of $|z'/z|^2$ are basically the phase space
ratios, which are
\begin{equation}
 \frac{|\vec q(Z_b^\prime\to h_b(1P)\pi)|^3}{|\vec q(Z_b\to h_b(1P)\pi)|^3}=1.20,
 \qquad  \frac{|\vec q(Z_b^\prime\to h_b(2P)\pi)|^3}{|\vec q(Z_b\to h_b(2P)\pi)|^3}=1.53.
\end{equation}
This implies that the ratios for the decays into the $h_b(mP)\pi$
are determined by the ratio of the partial decay widths for the
open-bottom decay modes,
\begin{equation}
\xi_m \simeq \frac{\text{PS}'_m}{\text{PS}_m} \frac{\Gamma(Z_b^{\prime+}\to B^{*+}\bar B^{*0})}
{\Gamma(Z_b^+\to B^+\bar B^0+B^0\bar B^{*+})},
\label{xiinterpret}
\end{equation}
where $\text{PS}_m^{(\prime)}$ is the phase space for the decays
$Z_b^{(\prime)}\to h_b(mP)\pi$.  Such a relation cannot be obtained were the
$Z_b$ states of tetraquark structure because the decay of a $\bar b\bar q b q$
tetraquark into the $h_b\pi$ knows nothing about the decay into the open-bottom
channels. Here, if we impose spin symmetry for $z$ and $z'$, one would get for
$\xi_m$ simply the ratio of phase spaces. In reality the second factor which is
the square of the ratio of effective couplings, c.f.
Eq.~(\ref{eq:coupling-ratio}), deviates from unity due to spin symmetry
violations enhanced by the proximity of the $B^{(*)}\bar B^{*}$ thresholds as
discussed above.

\begin{table}
\small \caption{The ratio $\Omega$ and the corresponding branching
fractions for all possible radiative decays. Uncertainties are given,
whenever they can be controlled theoretically (see text). The values
quoted without uncertainties may be understood as order of magnitude
estimates.}
\begin{center}
\begin{tabular}{|ccccc|}
               \hline\hline
                 & \multicolumn{2}{c}{$Z_b$} & \multicolumn{2}{c|}{$Z'_b$} \\
               & $\Omega$ & Branching Fraction & $\Omega$ & Branching Fraction  \\\hline
               $\chi_{b0}(1P)\gamma$ & $5\times10^{-3}$ & $1 \times10^{-4}$ & $4\times 10^{-3}$ & $3\times10^{-4}$ \\
               $\chi_{b1}(1P)\gamma$ & $1\times10^{-2}$ & $3 \times10^{-4}$ & $1\times 10^{-2}$ & $8\times10^{-4}$ \\
               $\chi_{b2}(1P)\gamma$ & $2\times10^{-2}$ & $ 5 \times10^{-4}$ & $2\times 10^{-2}$ & $1\times10^{-3}$ \\
               $\chi_{b0}(2P)\gamma$ & $(6.3\pm1.8)\times10^{-3}$ & $(2.7\pm1.5)\times10^{-4}$ & $(4.2\pm1.2)\times10^{-3}$ & $(6.2\pm 3.2)\times10^{-4}$ \\
               $\chi_{b1}(2P)\gamma$ & $(1.3\pm0.4)\times10^{-2}$ & $(5.6\pm 3.2)\times10^{-4}$ & $(1.3\pm0.4)\times10^{-2}$  & $(1.9\pm 1.0 )\times10^{-3}$ \\
               $\chi_{b2}(2P)\gamma$ & $(1.9\pm0.5)\times10^{-2}$ & $(8.3\pm 4.5)\times10^{-4}$ & $(1.8\pm0.5)\times10^{-2}$  & $(2.7\pm 1.3)\times10^{-3}$ \\
               \hline\hline
             \end{tabular}
\end{center}
\label{tab:radiative-decays}
\end{table}
Heavy quark spin symmetry allows one to gain more insight into the
molecular structure.  Because the $\chi_{bJ}(mP)$ are the
spin-multiplet partners of the $h_b(mP)$, the radiative decays of
the neutral $Z_b^{(\prime)0}$ into $\chi_{bJ}(mP)\gamma$ can be
related to the hadronic decays of the $Z_b^{(\prime)}$, no matter
whether they are
neutral or charged, into the $h_b(mP)\pi$. It is therefore useful to
define the following ratios
\begin{equation}
 \Omega_{\left[Z_b^{(\prime)},\chi_{bJ}(mP)\right]}:=
 \frac{\Gamma(Z_b^{(\prime)0}\to \chi_{bJ}(mP)\gamma)}{\Gamma(Z_b^{(\prime)0}\to h_b(mP)\pi^0)}.
\label{eq:ratio-decay}
\end{equation}
With the coupling constants in the bottom meson--photon Lagrangian
determined from  elsewhere, see for instance Ref.~\cite{Hu:2005gf},
such ratios can be predicted with no free parameters. At the
hadronic level, the $Z_b$ radiative transitions can only be related
to the hadronic ones if the $Z_b$'s are hadronic molecules so that
the two different types of transitions involve the same set of
coupling constants (modulo the bottom meson--pion/photon coupling
which can be determined from other processes or lattice
simulations). Thus, if the branching fractions of the radiative
transitions are large enough to be detected, such a measurement
would provide valuable information on the nature of the $Z_b$
states. The results for the ratios $\Omega$ are collected in
Tab.~\ref{tab:radiative-decays}.

Using the branching ratio $\mathcal{B}(Z_b^{({\prime})}\to h_b(1P,2P)\pi^+)$, we
find that the branching ratios of $Z_b^{(\prime)}\to \chi_{bJ}(1P,2P)\gamma$ are
of order $10^{-4}\sim 10^{-3}$. The largest branching fractions of
the $\chi_{bJ}(mP)$ are those into the $\gamma\Upsilon(nS)$, and the $\Upsilon(nS)$
can be easily measured. Thus, the final states for measuring the
$Z_b^0\to\chi_{bJ}\gamma$ would be the same as those of the
$Z_b^0\to\Upsilon(nS)\pi^0$ because the $\pi^0$ events are selected from photon
pairs. This means that the detection efficiency and background of these two
processes would be similar. 
In the preliminary experimental results~\cite{Adachi:2012im}, the $Z_b(10610)^0$
event number collected in the $\Upsilon(1S,2S)\pi^0$ channels is of order
$\mathcal{O}(100)$.   Given that the luminosity of the future
Super-KEKB could be two orders of magnitude higher than KEKB,
such transitions will hopefully be measured. Furthermore, one may also expect to
measure these radiative transitions at the LHCb. Note that the experimental
confirmation of the ratios given in Tab.~\ref{tab:radiative-decays} would be a
highly nontrivial evidence for the molecular nature of the $Z_b$ states.

Lacking knowledge of the $\chi_{bJ}(3P)B\bar B$ coupling constant,
the transitions into the $3P$ states cannot be predicted
parameter-free. However, they can be used to check the pattern in
Eq.~\eqref{eq:Aratio} predicted by the power counting analysis. The
decay widths of the $Z_b\to\chi_{bJ}(mP)\gamma$ are proportional to
$g_{1,mP}^2$. Taking the same value for $g_{1,mP}^2$, the explicit
evaluation of the triangle loops gives $ 1:1.8:4.4$ for the ratios
defined in Eq.~\eqref{eq:Aratio} with $J=1$. The values are close to
the ones in Eq.~\eqref{eq:Aratio}, and thus confirm the $1/\bar v$
scaling in Eq.~\eqref{eq:Apower} of the amplitudes.

\begin{table}[t]
\caption{The ratios defined in Eq.(\ref{eq:radiative-ratio}) for all
channels. Uncertainties are given, whenever they can be controlled
theoretically.}
\medskip
\centering
\begin{tabular}{|cc|}
               \hline\hline
                 & $(J=0):(J=1):(J=2)$ \\\hline
               $Z_b\to\chi_{bJ}(1P)\gamma$ & $1:2.5: 3.7$ \\
               $Z_b^\prime\to\chi_{bJ}(1P)\gamma$ & $1:2.9: 4.4$ \\
               $Z_b\to\chi_{bJ}(2P)\gamma$ & $1:(2.1\pm0.6):(2.9\pm 0.8)$ \\
               $Z_b^\prime\to\chi_{bJ}(2P)\gamma$ & $1:(3.0\pm 0.9): (4.2\pm 1.2) $ \\
               \hline\hline
             \end{tabular}
\label{tab:radiative-ratio}
\end{table}
As observed in Ref.~\cite{Guo:2010ak}, the NREFT leading loop calculation
preserves  the heavy quark spin structure. Because the $Z_b$ contains both
$s_{b\bar b}=0$ and 1 components, the leading contribution to its transitions
into the normal bottomonia, which are eigenstates of $s_{b\bar b}$, comply with
the spin symmetry. This conclusion should be true no matter what nature the
$Z_b$'s have as long as the spin structure does not change. Thus, one expects
that the branching fraction ratios of the decays of the same $Z_b$ into a
spin-multiplet bottomonia plus a pion or photon, such as
\begin{eqnarray}
\mathcal{B}(Z_b^{(\prime)}\to
\chi_{b0}(mP)\gamma): \mathcal{B}(Z_b^{(\prime)}\to\chi_{b1}(mP)\gamma): \mathcal{B}(Z_b^{(\prime)}\to\chi_{b2}(mP)\gamma),
\label{eq:radiative-ratio}
\end{eqnarray}
are insensitive to the structure of the $Z_b$. This statement may be confirmed
by observing that our results, as shown in Tab.~\ref{tab:radiative-ratio} agree
with the ratios $1:2.6:4.1$ (for the $1P$ states) and $1:2.5:3.8$ (for the $2P$
states) which are obtained solely based on heavy quark spin symmetry in
Ref.~\cite{Ohkoda:2012rj}. One may expect a derivation from the spin symmetry
results, which is due to the mass difference between the $B$ and $B^*$ mesons,
to be of order $\mathcal{O}(2\Lambda_{\rm QCD}/m_B)\sim 10\%$. The central
values given in Tab.~\ref{tab:radiative-ratio} deviate from the spin symmetry
results by at most 20\%, and they are fully consistent considering the
uncertainties.

\section{Comparison with other works on {\boldmath$Z_b$} decays}\label{sec:comparison}

Since their discovery in an impressive number of theoretical works  the
molecular nature of the $Z_b$ states was investigated. In this section we
compare in some detail our approach to the calculations in
Refs.~\cite{Mehen:2011yh,Dong:2012hc,livoloshin,Li:2012as} which deal with some
of the decays considered in our paper.
Common to most of these works is that, contrary to our approach, the  second
part of the one loop integral shown in Fig.~\ref{fig:PowerCounting1} (e) is
either approximated~\cite{Mehen:2011yh,Dong:2012hc} or  calculated differently~\cite{livoloshin}.
Especially, by this the analytic structure of the loop gets changed converting
it to a topology of type (f) in that figure, since
 the second $B^{(*)}\bar
B^{(*)}$ cut was removed from the loop. However, our power counting gives
that it is exactly this cut that drives the enhancement of the one loop diagrams
compared to the two-loop diagrams. 

Since the loop of type (f) gives the wave function
at the origin in $r$-space, the formalism applied in Refs.~\cite{Mehen:2011yh,Dong:2012hc,livoloshin} is  basically identical to that used in the classic calculations
for the decay of positronium into two photons.  However, as discussed in detail
in Ref.~\cite{scalardecay}, it is applicable only if the range of the transition potential
from the constituents to the final state is significantly shorter ranged than the 
potential that formed the molecule --- a scale not to be mixed up with the size
of the molecule which can be very large for a shallow bound state. 
However, the range of the binding momentum of the $Z_b$ states is not
known and might well be of the order of the range of the transition potential (at
least as long as the final bottomonium is not a ground state). In such a situation
in Ref.~\cite{scalardecay} it is proposed to calculate the full loop function for the
transitions, as done here in our work, which in effect means to expand around the
limit of a zero range potential that forms the bound state. In that paper it is also shown
that the potentially most important corrections to the transition rate cancel, such
that the uncertainty of the procedure is given by the binding momentum of the
molecule in units of the range of forces and not of the order of the final momenta
in units of the range of forces. This gives an additional justification for the
approach we are using.
We now discuss the formalisms of Refs.~\cite{Mehen:2011yh,Dong:2012hc,livoloshin,Li:2012as} 
in some more detail.

In Ref.~\cite{Mehen:2011yh} an effective field theory called X-EFT is
used. It is valid for hadronic
molecules with small binding energies so that the pion mass and the heavy meson
hyperfine splitting are hard scales. The decays of the $Z_b^{(\prime)}$ can be
represented by a bubble with two vertices, one connecting the $B^{(*)}\bar B^*$
to the $Z_b^{(\prime)}$ states and the other is a local operator for the
$B^{(*)}\bar B^*\pi(\bar b b)$ coupling. The coefficient of the local operator
depends on the pion energy, and is obtained by matching to the tree-level diagrams
in heavy hadron chiral perturbation theory.
For more details, we refer to
Refs.~\cite{Fleming:2007rp,Fleming:2008yn}. The ratios of $\Gamma(Z'\to
\Upsilon(3S)\pi)/\Gamma(Z\to \Upsilon(3S)\pi)$ and $\Gamma(Z'\to
h_b(2P)\pi)/\Gamma(Z\to h_b(2P)\pi)$ were calculated in Ref.~\cite{Mehen:2011yh}
assuming $z=z'$. As outlined in the discussion below Eq.~(\ref{xiinterpret}), 
in this case these ratios are just the ratios of phase spaces and thus
our results for them agree for $z=z'$.
The method of Ref.~\cite{Dong:2012hc} is a phenomenological variant of
the approach outlined above.

Also in Ref.~\cite{livoloshin} the transition
from the intermediate $B^{(*)}\bar B^{*}$-system to the final $\pi (\bar bb)$
system was assumed to be local, however, here the strength of this local
operator was calculated differently: the authors estimate it via the overlap
integral of the $\bar bb$ component in the $B^{(*)}\bar B^{*}$ wave function
with the outgoing $\bar bb$ pair in the presence of a dipole operator. While
this procedure is certainly justified when there are $1P$ states in the final
state
--- here the relative momenta between the two $B$ mesons are beyond 2.5 GeV and
indeed in this case our effective field theory does not converge anymore (c.f.
Sec.~\ref{sec:nonrel}) --- we regard it as questionable for the $2P$ states.
There is one more difference, namely the fact that in the formalism of
Ref.~\cite{livoloshin}, the transitions to the $\gamma \chi_{bJ}(nP)$ final
states are disconnected from those to the $\pi h_b(nP)$ states, while in our
approach they are connected as discussed in detail above. Thus, an experimental
observation of the decay of one of the $Z_b$ states to, say, $\gamma
\chi_{bJ}(2P)$ would allow one to decide on the applicability of our approach.

Similar to our work, in Ref.~\cite{Li:2012as} the full heavy meson loop is
evaluated, but regularized with a form factor. Absolute predictions are given
for the transitions using a model to estimate the $B^{(*)}\bar B^{(*)}(\bar bb)$
coupling --- it is difficult to judge the uncertainty induced by this. In a
first step in that work a cut-off parameter was adjusted to reproduce each
individual transition. It is found that the cut-off parameters needed for the
$h_b\pi$ transitions are typically larger and closer together than those needed
for the $\Upsilon\pi$ transitions. This hints at form factor effects being not
very significant in the former decays. This interpretation is also supported by
the observation that the ratios of decay rates --- the same quantities as
investigated here --- are found to be basically independent of the form factor.
In this sense the phenomenological studies of Ref.~\cite{Li:2012as} provide
 additional support for the effective field theory calculation presented here, 
 although a well-controlled error estimate cannot be expected from such a method.

\section{Summary}
\label{sec:summary} In this paper, we assume that both $Z_b$ and $Z_b^\prime$
are hadronic molecules predominantly coupling to $B\bar B^*$ and $B^*\bar B^*$,
respectively, in line with the data by the Belle Collaboration. As a consequence
of this assumption, the $Z_b$ states can only couple through $B^{(*)}\bar
B^*$--loops. Using NREFT power counting we argue that
\begin{itemize}
\item the decay channels $Z_b^{(\prime)}\to\Upsilon(nS)\pi$ as well as the
    transitions into the ground state $P$-wave bottomonia in the final state
    can not be controlled within the effective field theory, since higher loop
    contributions are expected to dominate the transitions;
\item model-independent predictions can be provided for $Z_b^{(\prime)}\to
    h_b(2P)\pi$ and radiative decays $Z_b^{(\prime)}\to
    \chi_{bJ}(2P)\gamma$.
\end{itemize}
The ratios for $Z_b$ and $Z'_b$ decays into the same final states
$h_b(mP)\pi$ are  consistent with the experimental data. Our results
reflect the fact that those ratios are essentially the ratio of the
corresponding phase space factors times the ratio of the $Z_bB\bar
B^*$ and $Z'_bB^*\bar B^*$ couplings squared. If further
experimental analysis with higher statistics could underpin this
fact, it would be a very strong evidence for the molecular
interpretation since such a relation cannot be obtained from, e.g.,
a tetraquark structure.

Furthermore, we calculate branching fractions for the final states
$\chi_{bJ}(mP)\gamma$. They are predicted to be of order
$10^{-4}\sim10^{-3}$. Although this is clearly a challenge to
experimentalists, a confirmation of these rates would strongly
support the molecular picture. It is noted that the ratios of a
certain $Z_b$ into bottomonia in the same spin multiplet are
insensitive to the structure of the $Z_b$, and may be obtained
solely based on heavy quark spin symmetry.

\section*{Acknowledgement}

We would like to thank M. Voloshin for valuable discussions. This work is supported
in part by the DFG and the NSFC through funds provided to
the Sino-German CRC 110 ``Symmetries and the Emergence of Structure in QCD'',
the EU I3HP ``Study of Strongly Interacting Matter'' under the Seventh Framework
Program of the EU, and the NSFC (Grant Nos. 11035006, 11121092 and 11165005).

\medskip

\begin{appendix}
\section{Nonrelativistic approach}\label{app:calnonrel}
\renewcommand{\theequation}{\thesection.\arabic{equation}}
\setcounter{equation}{0} The basic three-point loop function worked
out using dimensional regularization in $D=4$ is
\begin{eqnarray}
&& I(m_1,m_2,m_3,\vec{q})\non\\ &=& \frac{-i}{8}
\int\!\frac{d^Dl}{(2\pi)^D} \frac{1}{
\left[l^0-\frac{\vec{l}^2}{m_1}+i\epsilon\right]}
 \frac{1}{\left[l^0+b_{12}+\frac{\vec{ l}^2}{m_2}-i\epsilon\right]}
 \frac{1}{\left[l^0+b_{12}-b_{23}-\frac{(\vec{l}-\vec{
q})^2}{m_3}+i\epsilon\right] } \non\\
&=& \frac{\mu_{12}\mu_{23}}{16\pi} \frac{1}{\sqrt{a}} \left[
\tan^{-1}\left(\frac{c'-c}{2\sqrt{a(c-i\epsilon)}}\right)  + \tan^{-1}\left(\frac{2a+c-c'}{2\sqrt{a(c'-a-i\epsilon)}}\right)
\right],
\end{eqnarray}
where $m_i(i=1,2,3)$ are the masses of the particles in the loop,
$\mu_{ij}=m_im_j/(m_i+m_j)$ are the reduced masses, $b_{12} = m_1+m_2-M$,
$b_{23}=m_2+m_3+q^0-M$, with $M$ the mass of the initial particle, and
$$
a = \left(\frac{\mu_{23}}{m_3}\right)^2 \vec{ q}^{\, 2}, \quad c =
2\mu_{12}b_{12}, \quad c'=2\mu_{23}b_{23}+\frac{\mu_{23}}{m_3}\vec{ q}^{\, 2}.
$$
In terms of the loop function given above, the amplitudes for $Z_b^{+}$ and
$Z_b^{\prime+}$ decays into $h_b\pi^{+}$ are
\begin{align}
\Amp_{Z_b^{+}h_b} &= \frac{2\sqrt{2}g g_1 z_1}{F_\pi} \sqrt{M_{h_b}M_{Z_b}} \epsilon_{ijk}q^i
\varepsilon_{Z_b}^j \varepsilon_{h_b}^k  \left[I(M_B,M_{B^*},M_{B^*},\vec{q})
+I(M_{B^*},M_B,M_{B^*},\vec{q})\right], \label{Aeq:ahb1}\\
\intertext{and} %
\Amp_{Z_b^{\prime+}h_b} &= \frac{2\sqrt{2}g g_1 z_2}{F_\pi} \sqrt{M_{h_b}M_{Z_b'}}
\epsilon_{ijk}q^i \varepsilon_{Z_b'}^j \varepsilon_{h_b}^k \left[I(M_{B^*},M_{B^*},M_B,\vec{q})
+I(M_{B^*},M_{B^*},M_{B^*},\vec{q})\right], \label{Aeq:ahb2}
\end{align}
respectively. In all these amplitudes, both the neutral and charged
bottom and anti-bottom mesons have been taken into account. The
amplitudes for $Z_b^{(\prime)0}$ into $\chi_{bJ}\gamma$ read
\begin{align}
\Amp_{Z_b^{0}\chi_{b0}\gamma}= & -\sqrt{\frac23}i\beta e g_1 z\sqrt{M_{\chi_b}M_{Z_b}}\epsilon_{ijk}q^i\varepsilon_{Z_b}^j \varepsilon_{\gamma}^k \left[I(M_B,M_{B^*},M_{B^*},\vec{q})
-3\,I(M_{B^*},M_B,M_B,\vec{q})\right], \\
\Amp_{Z_b^{\prime0}\chi_{b0}\gamma}= & -2i\sqrt{\frac23}\beta e g_1 z'\sqrt{M_{\chi_b}M_{Z_b}}\epsilon_{ijk}q^i\varepsilon_{Z_b}^j \varepsilon_{\gamma}^k I(M_{B^*},M_{B^*},M_{B^*},\vec{q}), \\
\label{Aeq:achib0}
\Amp_{Z_b^{0}\chi_{b1}\gamma}= & ~2i\beta e g_1 z\sqrt{M_{\chi_b}M_{Z_b}} (q^i g^{jk}-q^k g^{ij})
\varepsilon_{Z_b}^i \varepsilon_{\gamma}^j\varepsilon_{\chi_b}^k  I(M_{B^*},M_B,M_{B^*},\vec{q})\\
\Amp_{Z_b^{\prime0}\chi_{b1}\gamma}= & -2i\beta e g_1 z\sqrt{M_{\chi_b}M_{Z_b}} (q^i g^{jk}-q^k g^{ij})
\varepsilon_{Z_b}^i \varepsilon_{\gamma}^j\varepsilon_{\chi_b}^k  I(M_{B^*},M_{B^*},M_B,\vec{q}), \label{Aeq:achib1}
\intertext{and} %
\Amp_{Z_b^{0}\chi_{b2}\gamma}= & ~ \sqrt2i\beta e g_1 z\sqrt{M_{\chi_b}M_{Z_b}}q^i
(g^{jm}\varepsilon_{ikl}+g^{jl}\varepsilon_{ikm})\varepsilon_{Z_b}^j \varepsilon_{\gamma}^k
\varepsilon_{\chi_b}^{lm}  I(M_{B^*},M_{B^*},M_{B^*},\vec{q}) \label{Aeq:achib2},\\\nonumber
\Amp_{Z_b^{0\prime}\chi_{b2}\gamma}= & ~ \sqrt2i\beta e g_1 z\sqrt{M_{\chi_b}M_{Z_b}}
\left(q^i(g^{jm}\varepsilon_{ikl}+g^{jl}\varepsilon_{ikm})+q^l\varepsilon^{ijm}+q^m\varepsilon^{ijl}\right)\\
& \times\varepsilon_{Z_b}^j \varepsilon_{\gamma}^k
\varepsilon_{\chi_b}^{lm}  I(M_{B^*},M_{B^*},M_{B^*},\vec{q}),
\end{align}
respectively.

\section{Relativistic approach}\label{app:calrel}

In this appendix, we formulate a Lorentz covariant framework for a parallel
study of the $Z_b^{(')}$ decays. In such a framework, more terms will appear as
relativistic corrections that are generally neglected in the heavy quark limit.
Therefore, a comparison between these two prescriptions will serve as a
cross-check of the NREFT results, and more importantly as a confirmation of the
validity of the NREFT power counting.

The bottom meson fields are defined as
\begin{eqnarray}
\nonumber
H_1&=&\left(\frac{1+\rlap{/}{v}}{2} \right)\left[P^*_\mu\gamma^\mu-\gamma_5 P\right],
\quad H_1^\dag=\gamma^0\left[P^{*\dag}_\mu\gamma^\mu+\gamma_5 P^\dag\right]\left(\frac{1+\rlap{/}{v}}{2} \right)\gamma^0,
\quad \bar{H}_1=\gamma^0 H_1^\dag\gamma^0,\\\nonumber
H_2&=&\left[\bar{P}^*_\mu\gamma^\mu-\gamma_5 \bar{P}\right]\left(\frac{1-\rlap{/}{v}}{2} \right),
\quad H_2^\dag=\gamma^0\left(\frac{1-\rlap{/}{v}}{2} \right)\left[\bar{P}^{*\dag}_\mu\gamma^\mu+\gamma_5 \bar{P}^\dag\right]\gamma^0,
\quad \bar{H}_2=\gamma^0 H_2^\dag\gamma^0,
\label{eq-bottomed-field}
\end{eqnarray}
where $P^*$ and $P$ represent the $(B^{*+}, B^{*0}, B_s^{*0})$ and
$(B^{+}, B^{0}, B_s^{0})$ fields, respectively, which annihilate the
corresponding particles, while $\bar P^{*}$ and $\bar P$ are the
fields of their antiparticles. The fields annihilating the $S$- and
$P$-wave bottomonia are given by
\begin{eqnarray}
\nonumber
R_{b\bar{b}}&=&\left(\frac{1+\rlap{/}{v}}{2} \right)\left(\Upsilon^\mu\gamma_\mu-\eta_b\gamma_5\right)
\left(\frac{1-\rlap{/}{v}}{2} \right),\\\nonumber
P_{b\bar{b}}^\mu &=&\left(\frac{1+\rlap{/}{v}}{2} \right)\left[\chi_{b2}^{\mu\alpha}\gamma_\alpha
+ \frac{1}{\sqrt{2}}\epsilon^{\mu\nu\alpha\beta} v_\alpha\gamma_\beta\chi_{b1\nu}
+\frac{1}{\sqrt{3}}(\gamma^\mu-v^\mu)\chi_{b0}+h_b^\mu\gamma_5\right]\left(\frac{1-\rlap{/}{v}}{2} \right),\label{eq-P-wave}
\end{eqnarray}
respectively. The Lagrangians for $S$-wave ($P$-wave) quarkonia and
a pair of heavy mesons are
\begin{eqnarray}
\mathcal{L}_{SB\bar{B}}&=&ig_2\mathrm{Tr}\left[R_{b\bar{b}}\bar{H}_{2a}\gamma^\mu
\overleftrightarrow{\partial}_\mu\bar{H}_{1a}\right]+H.c.~,\label{eq:upsilon}\\
\mathcal{L}_{PB\bar{B}}&=&ig_1\mathrm{Tr}\left[P^\mu_{b\bar{b}}\bar{H}_{2a}\gamma_\mu
\bar{H}_{1a}\right]+H.c. \ \label{eq:hb}
\end{eqnarray}
Under the similar convention, the $Z_b$ field can be expressed as
\begin{eqnarray}
P_Z^\mu=\left(\frac{1+\rlap{/}{v}}{2} \right)Z^\mu\gamma_5\left(
\frac{1-\rlap{/}{v}}{2} \right)~,\label{eq-Zb}
\end{eqnarray}
and the effective interaction between the $Z_b$ and a pair of bottomed meson reads
\begin{eqnarray}
\mathcal{L}_{ZB\bar{B}}&=&iz^{(\prime)}\mathrm{Tr}
\left[P^{\dag\mu}_{Z,ab}\bar{H}_{2b}\gamma_\mu\bar{H}_{1a}\right]+H.c.~,
\label{eq-ZBB}
\end{eqnarray}
The Lagrangian for the pion coupling to a pair of bottom mesons
is~\cite{Burdman:1992gh,Wise:1992hn,Yan:1992gz}
\begin{eqnarray}
\mathcal{L}=ig\mathrm{Tr}[H_b\gamma_\mu\gamma_5\mathcal{A}_{ba}^\mu\bar{H}_a]
\ ,
\end{eqnarray}
where $\mathcal{A}_\mu=(\xi^\dag\partial_\mu\xi-\xi\partial_\mu\xi^\dag)/2$,
with $\xi=\mathrm{exp}(i\sqrt{2}\phi/F_\pi)$. The relativistic form of the
Lagrangian for the photon coupling to the bottom mesons is~\cite{Hu:2005gf}
\begin{eqnarray}
\mathcal{L}_\gamma&=&\frac{e\beta Q_{ab}}{2}F^{\mu\nu}\mathrm{Tr}[H_b^\dag \sigma_{\mu\nu}H_a]
+\frac{eQ^\prime}{2m_{Q}}F^{\mu\nu}\mathrm{Tr}[H_a^\dag H_a\sigma_{\mu\nu}].
\end{eqnarray}
With these effective Lagrangians, we obtain for  the amplitudes of $Z_b^{(\prime)}\to h_b\pi$
\begin{eqnarray}
\mathcal{A}_{Z_bh_b}&=&\frac{2\sqrt{2}gg_1z}{F_\pi}\epsilon_{q v \epsilon_{h_b}
\epsilon_{Z_b}}(C_0[M_B,M_{\bar{B}^*},M_{B^*}]+C_0[M_{B^*},M_{\bar{B}},M_{B^*}])~,\\
\mathcal{A}_{Z_b^\prime h_b}&=&\frac{2\sqrt{2}gg_1z}{F_\pi}\epsilon_{q v \epsilon_{h_b}
\epsilon_{\epsilon_{Z_b}}}(C_0[M_{B^*},M_{\bar{B}^*},M_{B^*}]+C_0[M_{B^*},M_{\bar{B}^*},M_{B}]).
\end{eqnarray}
and to $\Upsilon\pi$
\begin{eqnarray}\label{A-Zb-upsilon-pi}
\nonumber
\mathcal{A}_{Z_b\Upsilon}&=&\frac{2\sqrt{2}gg_2z}{F_\pi}(-\epsilon_\Upsilon\cdot
\epsilon_Z|
\vec{q}|^2(C_0[M_{B},M_{\bar{B}^*},M_{B^*}]+2C_2[M_{B},M_{\bar{B}^*},M_{B^*}]\\\nonumber
&+&C_0[M_{B^*},M_{\bar{B}},M_{B^*}]+2C_2[M_{B^*},M_{\bar{B}},M_{B^*}])+q\cdot
\epsilon_\Upsilon q\cdot \epsilon_Z(C_0[M_{B^*},M_{\bar{B}},M_{B}]\\
&+&2C_2[M_{B^*},M_{\bar{B}},M_{B}]-C_0[M_{B^*},M_{\bar{B}},M_{B^*}]
-2C_2[M_{B^*},M_{\bar{B}},M_{B^*}]))~,
\end{eqnarray}
\begin{eqnarray}\label{A-Zb'-upsilon-pi}\nonumber
\mathcal{A}_{Z_b^\prime \Upsilon}&=&\frac{2\sqrt{2}gg_2z}{F_\pi}
(\epsilon_Z\cdot\epsilon_\Upsilon|\vec{q}|^2(C_0[M_{B^*},M_{\bar{B^*}},M_B]+C_2[M_{B^*},M_{\bar{B^*}},M_B]\\\nonumber
&+&C_0[M_{B^*},M_{\bar{B^*}},M_{B^*}]+C_2[M_{B^*},M_{\bar{B^*}},M_{B^*}])
+q\cdot \epsilon_\Upsilon q\cdot \epsilon_Z(C_0[M_{B^*},M_{\bar{B^*}},M_B]\\
&+&C_2[M_{B^*},M_{\bar{B^*}},M_B]-C_0[M_{B^*},M_{\bar{B^*}},M_{B^*}]
-C_2[M_{B^*},M_{\bar{B^*}},M_{B^*}]))~.
\end{eqnarray}
The amplitudes for $Z_b^{(\prime)}\to \chi_{b0}\gamma$ are
\begin{eqnarray}
A_{Z_b^0\chi_{b0}\gamma}&=&i\frac{\beta eg_1z\sqrt{2}}{\sqrt{3}}\epsilon_{q v \epsilon_{\gamma} \epsilon_{Z_b}}
(3C_0[M_{B^*},M_{\bar{B}},M_{B}]-C_0[M_{B},M_{\bar{B^*}},M_{B^*}])\\
A_{Z_b^0\chi_{b1}\gamma}&=&i2eg_1z\beta(\epsilon_{\chi_{b1}}\cdot\epsilon_\gamma
q\cdot\epsilon_Z-q\cdot \epsilon_{\chi_{b1}}\epsilon_{\gamma}\cdot \epsilon_{Z} )C_0[M_{B^*},M_{\bar{B}},M_{B^*}]\\
A_{Z_b^0\chi_{b2}\gamma}&=&i2\sqrt{2}eg_1z\beta\epsilon_{Z}^\sigma
\epsilon^{\alpha q v \epsilon_{\gamma}}\epsilon_{\chi_{b2}\sigma\alpha}
C_0[M_{B},M_{\bar{B^*}},M_{B^*}]~,
\end{eqnarray}
and
\begin{eqnarray}
A_{Z_b^{\prime0}\chi_{b0}\gamma}&=&-2\frac{\sqrt{2}}{\sqrt{3}}ieg_1z^\prime \beta
\epsilon_{qv\epsilon_{\gamma}\epsilon_Z}C_0[M_{B^*},M_{\bar{B^*}},M_{B^*}] \\
A_{Z_b^{\prime0}\chi_{b1}\gamma}&=&-i2eg_1z^\prime\beta (\epsilon_{\chi_{b1}}\cdot \epsilon_{\gamma}
q\cdot \epsilon_Z-q\cdot \epsilon_{\chi_{b1} }\epsilon_{\gamma}\cdot \epsilon_{Z}) C_0[M_{B^*},M_{\bar{B^*}},M_{B^*}]\\
A_{Z_b^{\prime0}\chi_{b2}\gamma}&=&i2\sqrt{2}e g_1 z^\prime \beta (q^\alpha
\epsilon_{\sigma v \epsilon_\gamma \epsilon_Z}+\epsilon_\gamma^\alpha\epsilon_{\sigma q v \epsilon_Z})\chi_{b2\alpha}^\sigma C_0[M_{B^*},M_{\bar{B^*}},M_{B^*}]~,
\end{eqnarray}
respectively. Here, $C_0$ is the standard relativistic three-point scalar
function which is similar to the definition of $I$ in App.~\ref{app:calnonrel}
\begin{eqnarray}\nonumber
&&C_0[p^2,(-q)^2,(p-q)^2,m_2^2,m_1^2,m_3^2]\\
&=&\frac{(2\pi\mu)^{4-D}}{i\pi^2}\int \frac{d^D l}{(l^2-m_1^2)((p-l)^2-m_2^2)((l-q)^2-m_3^2)}~,
\end{eqnarray}
where the incoming four-momentum of the  $Z_b$ is $p$,
 and the light outgoing particle four--momentum  is $q$.
For simplicity, we do not explicitly include the first three arguments in $C_0$
in the above formulaes. With the same convention, $C_2$ is a coefficient that
arises from the tensor reduction of the three-point vector loop,
\begin{eqnarray}\nonumber
C^\mu&\equiv&\frac{(2\pi\mu)^{4-D}}{i\pi^2}\int \frac{l^\mu d^D l }{(l^2-m_1^2)((p-l)^2-m_2^2)((l-q)^2-m_3^2)}\\
&=&C_1p^\mu+C_2(p-q)^\mu.
\end{eqnarray}
which is similar to the integral $I_1$ defined in Ref.~\cite{Guo:2010ak}.

To compare with the NREFT method, we note the following treatments in the heavy
quark limit:
\begin{itemize}
\item By expressing the heavy meson momentum $M_Q v_\mu=m_Qv_\mu+k_\mu$ in the
    heavy quark limit, where $v$ is the heavy meson velocity and $k$ is the
    residual momentum of the order of $\Lambda_{QCD}$, the derivative in
    Eq.~(\ref{eq:upsilon}) only gives the difference of the residual momentum
    between the intermediate bottomed mesons~\cite{Colangelo:2003sa}. Namely,
    the residual momentum $k$ is the integral momentum in the meson loops
    instead of $v$.
\item The denominators $(l^2-m_1^2)((p-l)^2-m_2^2)((l-q)^2-m_3^2)$ of the
    meson loops contain two independent external momenta $p$ and $q$. For the
    incoming momentum $p$, we set $p=M_Zv$ following the convention of
    Ref.~\cite{Guo:2010ak}, where $v=(1,0,0,0)$ defines the rest-frame of the
    initial particle. By doing this, the vector three-point function defined
    in the loop integrals becomes $C^\mu=M_Zv^\mu C_1+(M_Z v^\mu-q^\mu)C_2$
    which can be compared with the NREFT amplitudes term by term. It is
    interesting to note that the coefficient of $C_1$ vanishes in the decay
    amplitude because of cancellation.

\item Another difference between this scheme and the NREFT comes from the
    contraction term $q\cdot \epsilon_\Upsilon$, where $q$ is the external
    light meson (e.g. pion) momentum. Note that the time component appears as
    an additional contribution compared to the NREFT formalism. It is
    proportional to $|\vec{q}|E_\pi/m_\Upsilon$, which is a relativistic
    correction and relatively suppressed with respect to the space component
    $|\vec{q}|E_\Upsilon/m_\Upsilon$.
\item The relativistic corrections  also arise from the mass difference
    between $m_B$ and $m_{B^*}$. One notices that in
    Eqs.~(\ref{A-Zb-upsilon-pi}) and (\ref{A-Zb'-upsilon-pi}), the terms
    proportional to $q\cdot \epsilon_\Upsilon q\cdot \epsilon_Z$ are given by
    the $D$-wave transition. In the heavy quark limit with $m_B=m_{B^*}$,
    exact cancelations occur within the integral functions $C_0$ and $C_2$,
    respectively. This means if the $Z_b$ and $Z_b^\prime$ are indeed the
    $B\bar B^*$ and $B^*\bar B^*$ molecular states, respectively, their
    $D$-wave decays into the $\Upsilon\pi$ will be highly suppressed. This can
    be understood by noticing that the heavy quark spin decouples from the
    system in the heavy quark limit, and the total angular momentum of the
    light quarks $s_{q\bar q}$ is a good quantum number. If the
    $Z_b^{(\prime)}$ states are $S$-wave $B^{(*)}\bar B^*$ hadronic
    molecules as assumed here, there is no spatial angular momentum in the
    system. Thus, the light quark system has $s_{q\bar q}=0$ or 1, and
    therefore cannot couple to a spinless pion in a $D$-wave. A similar
    statement was made very recently in Ref.~\cite{Voloshin:2013ez}. Here we
    notice that the decay of a $b\bar bq\bar q$ tetraquark state with
    $s_{q\bar q}^P=2^-$ would decay into $b\bar b\pi$ dominantly in a
    $D$-wave.

\item We also note the convention for the sums of the polarizations
for the vector and tensor particles:
      $\epsilon_{\Upsilon(h_b,\chi_{b_1})}^\mu\epsilon^{*\nu}_{\Upsilon(h_b,\chi_{b_1})}=-g^{\mu\nu}+v^\mu
      v^\nu\equiv \widetilde{g}^{\mu\nu}$ and
      $\epsilon_{\chi_{b2}}^{\mu\nu}\epsilon_{\chi_{b2}}^{*\alpha\beta}=\frac{1}{2}(\widetilde{g}^{\mu\alpha}\widetilde{g}^{\nu\beta}
      +\widetilde{g}^{\mu\beta}\widetilde{g}^{\nu\alpha})-\frac{1}{3}\widetilde{g}^{\mu\nu}\widetilde{g}^{\alpha\beta}$.
They allow us to separate out the relativistic contributions in the
scalar loop diagrams which can then be compared with the NREFT
formulaes explicitly.

\end{itemize}

The relativistic formalism is advantageous for providing a cross check of the
NREFT results and singling out the effects arising from the relativistic
corrections. For the bottomonium and bottomed meson system discussed here, it
shows that the relativistic corrections are indeed small and these two methods
are consistent with each other. Alternatively, it should be cautioned that in
the relativistic formalism, when the terms proportional to
$|\vec{q}|E_\pi/m_{Q\bar{Q}}$ are not obviously suppressed in comparison with
the three-momentum $|\vec{q}|$, i.e. $E_\pi/m_{Q\bar{Q}}$ is sizeable, a
different scheme with a form factor~\cite{Wang:2011yh,Wang:2012wj} might be used
to control the large momentum contributions and non-local effects of each
couplings.

\end{appendix}

 \end{document}